\documentclass[11pt,twoside,a4paper]{article}

 \usepackage{amsmath}
 \usepackage{amssymb}
 \usepackage{epsfig}
 \usepackage{graphicx}
 \usepackage{psfrag}
 \usepackage[english]{babel}
 \usepackage{t1enc}
 \usepackage[latin9]{inputenc}
 \usepackage{xspace}
 \usepackage[left=2cm,right=2cm,top=2.5cm,bottom=2.5cm]{geometry}

 \title{}
 \date{}

\begin{document}
 \vspace{1.cm}
 \flushright{ULB-TH/08-01}
 \vspace{1.cm}

 \begin{center}

 {\Large {\bf  Gamma ray and Neutrinos fluxes from a cosmological dark matter simulation }}
 \vspace{0.8cm}\\
 {\large  E. Athanassoula$^1$, F.-S. Ling$^2$, E. Nezri$^{1,2}$, R.
 Teyssier$^3$}
 \vspace{0.35cm}\\

 $^1${\it Laboratoire d'Astrophysique de Marseille,\\ 
Observatoire Astronomique de Marseille Provence\\
CNRS/Universit\'e de Provence\\2 Place Le Verrier, 13248 Marseille
 C\'edex 04, France}
 \vspace{0.35cm}

 $^2${\it
  Service de Physique Th\'eorique, Universit\'e Libre de Bruxelles,\\
  CP225, Bld du Triomphe, 1050 Brussels, Belgium}
 \vspace{0.35cm}

 $^3${\it DAPNIA, CEA Saclay \\
 L'Orme des merisiers 91191 Gif-sur-Yvette, France}

 \vspace{0.5cm}

 \end{center}
 \abstract{ In this paper, we estimate the gamma-ray and neutrino fluxes coming from dark
matter annihilation in a Milky Way framework 
 provided by a recent N-BODY HORIZON simulation. We first study the characteristics of the simulation and
highlight the mass distribution within the galactic halo. The general dark matter density has a typical $r^{-3}$
power law for large radii, but the inner behaviour is poorly constrained
below the resolution of the simulation ($\sim 200$ pc). We identify
clumps and subclumps and analyze their distribution, as well as their internal
structure. Inside the clumps, the power law is rather universal, $r^{-2.5}$ in the outer part
with again strong uncertainties for smaller radii, especially for light
clumps.
We show a full-sky map of the astrophysical contribution to the
gamma-ray or neutrino fluxes in this N-body framework.
Using quite model independent and general assumptions for the
high energy physics part, we evaluate the possible absolute fluxes and
show some benchmark regions for the experiments GLAST, EGRET, and a
km3 size extension of ANTARES like the KM3NeT project. 
While individual clumps seem to be beyond detection reach, the galactic center
region is promising and GLAST could be sensitive to the geometry and the
structure of its dark matter distribution. The detection by a km3 version of ANTARES is, however, more challenging
due to a higher energy threshold.
We also point out that the lack of resolution leaves the inner structure of subhalos poorly constrained.
Using the same clump spectrum and mass fraction, a clump luminosity boost of order ten
can be achieved with a steeper profile in the inner part of the sub-halos.}

 \vspace{1.cm}

 \section{Introduction}

 An ever increasing number of
 observational and theoretical results strongly suggest, or even require 
 the existence of dark matter, whose enigma becomes thus crucial for the
 understanding of our universe.  Let us mention amongst others the
 WMAP results on 
 CMB \cite{Spergel:2006hy}, the rotation curves of disk galaxies
 \cite{Bosma:2003yv}, the formation of large 
 scale structures \cite{White:1987yr}, the
 bullet cluster observation \cite{Clowe:2003tk}, merger modeling and
 lensing results {\it e.g.} \cite{Ferreras:2007na}. Finally, the
 possibility of 
 numerous extensions of high energy physics beyond the standard model (BSM)
 to provide new weakly interacting massive particle (WIMP) candidates for
 dark matter makes the hypothesis very appealing.

 Nevertheless, the nature, the identification and the distribution of the
 dark matter are still open questions,
 intimately linked with the proof of its existence. Present and near
 future instrumental projects could bring welcome input to these questions.
 Namely, both astroparticle physics and collider searches will reach higher
 sensitivities with experiments like LHC (accelerator); EDELWEISS II,
 superCDMS and ZEPLIN (direct
 detection); ANTARES and ICECUBE (neutrino telescopes); GLAST (gamma
 space telescope) and PAMELA (charged particle search satellite). Amongst
 the different possibilities, indirect detection is particularly
 promising. Indeed, relic dark matter particles can accumulate in
 cosmic storage rings and annihilate. The decay of
 their annihilation products will give rise to secondary particle
 fluxes ($\gamma, \nu,e^+,\bar{p}$), which could be detected by
 dedicated experiments indirectly indicating the presence of dark matter.
 
 In this paper, we will focus on indirect detection of dark matter
 through gamma rays and neutrinos. Galaxies are thought to be interesting sources for
 this kind of detection, seen the amount of dark matter they are
 believed to harbour. As we will discuss later, the detectability of
 such gamma rays or neutrinos depends strongly on both the astrophysical assumptions
 on the dark matter distribution in the  halo and on the assumed high
 energy physics BSM scenario. Some studies
 concerning different particle physics models can be found in the
 literature (see \cite{Jungman:1995df,Bertone:2004pz} for reviews). The popular
 BSM dark matter scenarios are typically supersymmetric
 models, models with extra dimensions, light dark matter, little Higgs
 model, inert doublet model ... or any extensions providing WIMP. The
 Milky Way astrophysical framework is
 commonly simplified with assumptions of spherical symmetry, now known
 to be incorrect
 \cite{Athanassoula:2003yw,Allgood:2006,Athanassoula:2007ex,Heller:2007tr},
 and typical smooth dark matter density functions extracted from N-body
 simulations \cite{Navarro:1996gj,Kravtsov:1997dp,Moore:1999gc}. Few
 recent works \cite{Stoehr:2003hf,Diemand:2006ik,Kuhlen:2007wv} and
 especially \cite{Kuhlen:2008aw} with an impressive resolution treat
 in detail the astrophysical aspects of the gamma ray fluxes coming
 from dark matter annihilation in realistic simulation
 frameworks. Other works {\it e.g} \cite{Tasitsiomi:2002vh,Pieri:2003cq,Koushiappas:2003bn,Bi:2005im,Pieri:2007ir} consider also more sophisticated parametrization inspired by extrapolations of
 simulation results. 
 
 Typical simulation canvas consist of 1-100
 millions of particles with mass around $10^5-10^6$ solar masses. The
 results reproduce well the large-scale structure formation and have
 now shown that virialized systems  are still left with surviving
 subhalos, also called {\it  clumps} . The
 results are more and more promising with computing upgrades and resolution
 improvements. Nevertheless, some questions
 are still open with regard to observations. For instance, the
 radial density profiles predicted for the innermost region of
 galactic halos are 
 quite cuspy, whereas observations suggest flat cores (see
 \cite{Bosma:2003yv} for a review). Furthermore, simulations predict
 more numerous galactic satellites than observed for the Milky
 Way. Even if  N-body calculations may generate too concentrated objects,
 the simulated haloes are the only realistic or advanced dark matter
 distribution framework. Specifically, the estimation of dark matter
 detectability in
 our neighborhood depends on both the dark matter distribution in the Milky
 Way -- especially in the innermost region -- and on the number, the size and the
 concentration of the clumps. Depending on the assumptions or results on
 these key points, different results have been proposed in previous works
 \cite{Koushiappas:2003bn,Evans:2003sc,Pieri:2005pg,Oda:2005nv,Koushiappas:2006qq,Bi:2005im,Baltz:2006sv,Strigari:2006rd,Pieri:2007ir,Strigari:2007at}.

 The present article is devoted mainly to the astrophysical contribution
 concerning dark matter gamma and neutrino indirect detection. We calculated the
 possible gamma and neutrino fluxes sky map in a N-body simulation framework provided by
 a HORIZON project simulation \cite{horizonweb}. The paper is organized as follows:
 section 2 gives the analysis of the numerical simulation and highlights
 the resulting dark matter distribution. In section 3, the gamma ray
 and neutrino flux
 calculation is shortly reviewed and a comparison of our estimates with
 regard to GLAST and ANTARES reach is presented. Conclusion and perspectives
 are given in section 4.

 \section{Simulation characteristics}

\subsection{General features}

\begin{figure}
 \begin{center}
 \includegraphics[width=0.8\textwidth]{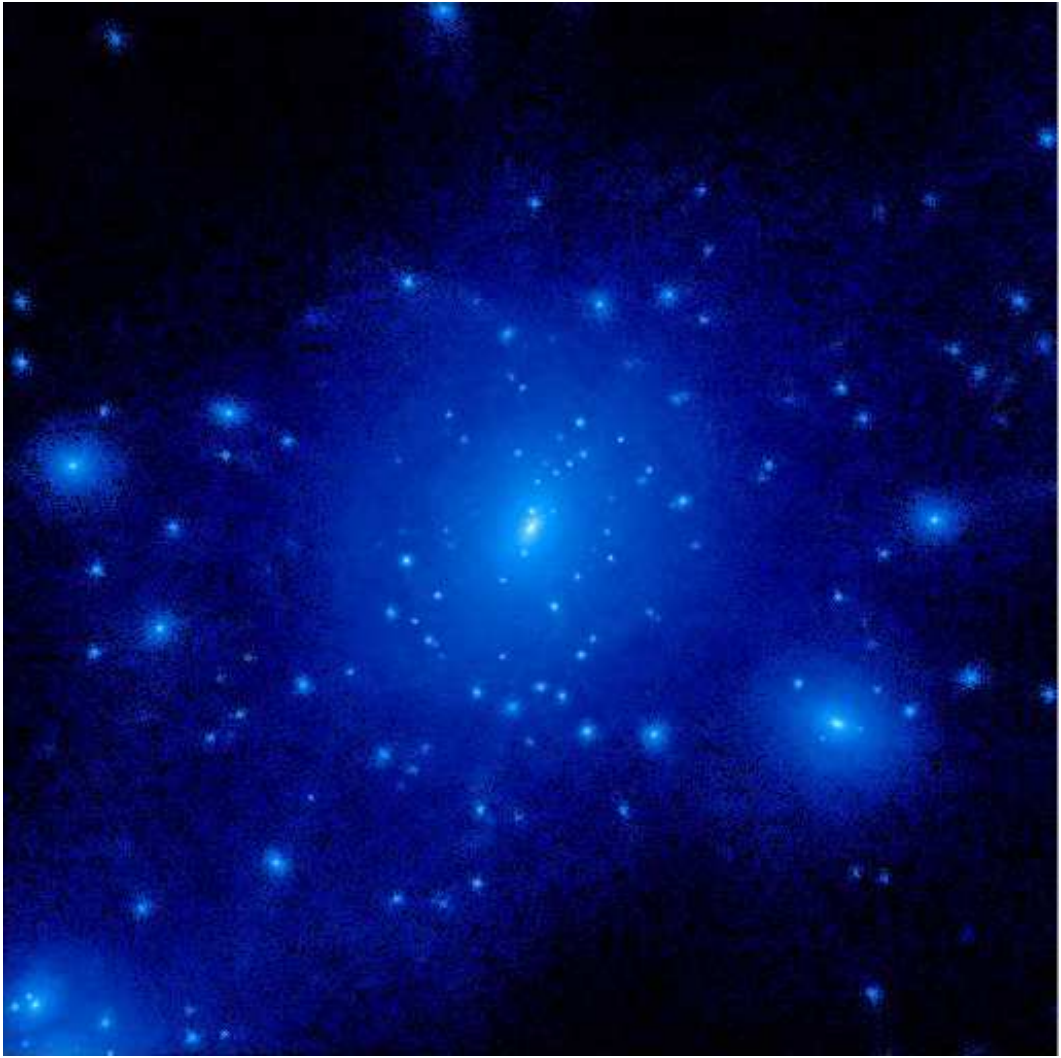}
 \caption{\small View of the galactic dark matter halo in the HORIZON simulation (projected along the z axis). 
 The box size corresponds to $750$ kpc.}
 \label{fig:simu}
 \end{center}
 \end{figure}

The  data  used   for  this  paper  were  provided   by  the  Horizon
collaboration.  The simulation was  performed using the Adaptive Mesh
Refinement code  RAMSES \cite{Ramses}.  The  initial conditions
are set by the WMAP3 results ($\Omega_m=0.24$, $\Omega_\Lambda=0.76$,
$\Omega_b=0.042$, $n=0.958$, $H_0=73$, $\sigma_8=0.77$) and
the effective  number of particles is  $N_p=1024^3$ in a  box of size
$L=20h^{-1}$  Mpc.  At $z=0$ we selected a 
Milky-Way sized halo.  Using the so--called
``zoom''  technique,  we  re-defined  the  grid outside  a  sphere  of
diameter $5h^{-1}$,  using high--mass  particles to sample  the large
scale tidal  field and smaller ones for the selected halo region. In
the  high--resolution region, we  increase the 
resolution of the grid on a cell-by-cell basis, with a maximum of $7$
additional levels  of refinement,  corresponding to a  maximum linear
resolution of about  $200$ pc. Cells are refined  if the local number
of dark  matter particles exceeds  10. Our smallest particle  mass is
$M_p=7.46\ 10^5$ in solar mass ($M_{sun}$) units.
Our simulated galactic environment is depicted in Fig.~\ref{fig:simu}.


 The size of a galactic halo is characterized by its virial radius, $r_{vir}$,
often defined as the size of the sphere centered on the galaxy center with an average density equal to  $200$
 times the cosmological matter density.
 In our simulation, the virial radius is equal to 253 kpc,
 corresponding to an enclosed mass of  $6.05 \times 10^{11} {\rm
   M_{\odot}}$ or $8.1\times 10^5$ particles.

Dark matter halos are often parameterized by spherically symmetric
profiles of the form
\begin{equation}
  \rho(r) = \rho_{0} ~\left(\frac{r}{r_0}\right)^{-\gamma} \left[
  \frac{1+\left(r_0/a\right)^{\alpha}}
  {1+\left(r/a\right)^{\alpha}}\right]^{\frac{\beta-\gamma}{\alpha}},
 \end{equation}
where  $\rho_0$ is the local density in the solar neighborhood,
$r_0=8$ kpc is the distance from the sun to the galactic centre, $\gamma$ is
the inner slope,  $\beta$ is the outer slope.  $\alpha$
describes the transition behavior around $r=a$.
The popular NFW profile  \cite{Navarro:1996gj} has $\alpha=1$, $\beta=3$
and $\gamma=1$ and, adapted to the Milky Way, corresponds to
, $a=20$ kpc and  $r_0=8$ kpc.

 \begin{figure}[t]
 \begin{center}
 \includegraphics[width=0.6\textwidth]{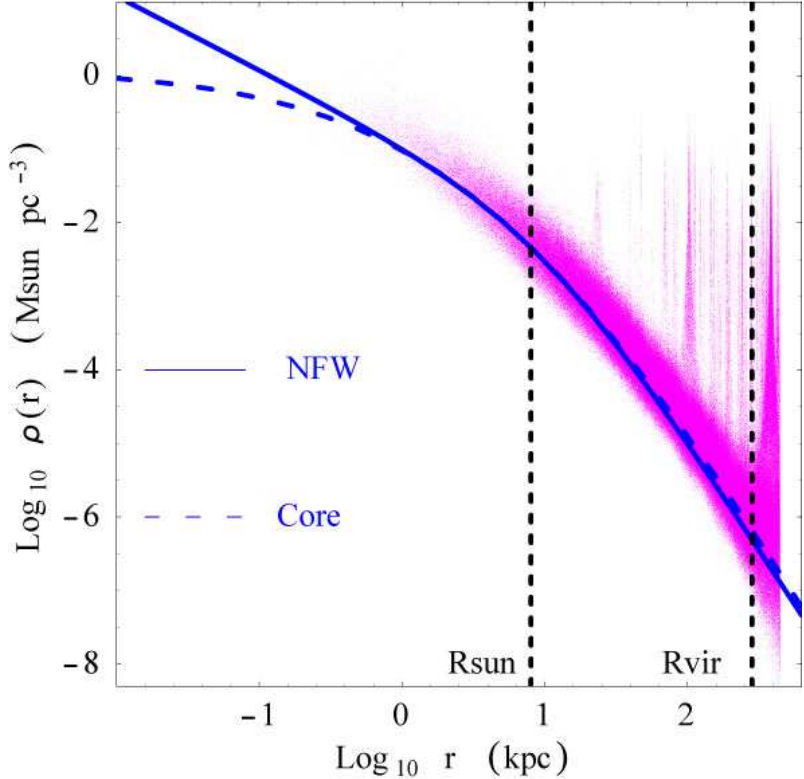}
 \caption{\small Density at all the positions of the particles in the
 simulation as a function of their distance to the center. This is
 compared to NFW ($\alpha=1$, $\beta=3$ and 
 $\gamma=1$, $a=10$ kpc, solid line) and core ($\alpha=0.5$, $\beta=3.3$, $\gamma=0$, $a=4.5$ kpc, dashed line) profiles.}
 \label{fig:rhoDM}
 \end{center}
 \end{figure}

  \begin{figure}[t]
 \begin{center}
 \includegraphics[width=0.6\textwidth]{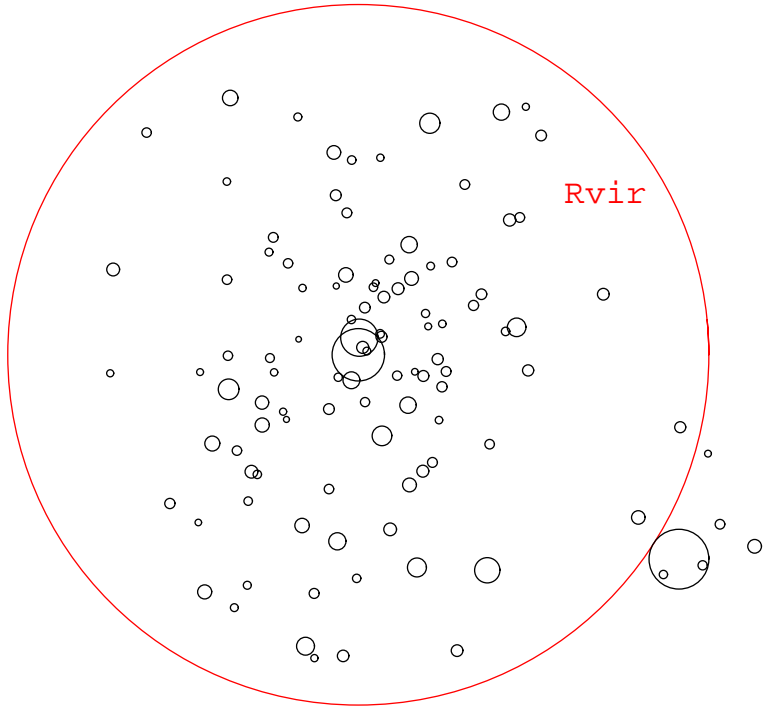}\\
 \caption{\small  Clump distribution in the galaxy, projected along the $z$ axis. 
 The location of each clump is identified by a circle, whose size scales as $M^{1/3}$. The red (largest) circle shows the virial radius
 (as defined in the text) of the galaxy (253 kpc).}
 \label{fig:clumpcircle}
 \end{center}
 \end{figure}

Fig.~\ref{fig:rhoDM} depicts the radial distribution of the
density for all particles in our simulation, calculated by the method
of Casertano and Hut~\cite{1985ApJ29880C}, as will be discussed
below. The center is taken as  
the densest point of the galactic halo. Due to resolution limits,
this distribution is globally
 consistent both with a NFW-like profile ($\alpha=1$, $\beta=3$ and
 $\gamma=1$, $a=10$ kpc) and a cored profile ($\alpha=0.5$,
 $\beta=3.3$, $\gamma=0$, $a=4.5$ kpc), with $r_0 = R_{sun}=8$ kpc in both
 cases. 
At any given radius, the densities exhibit a large spread due to non
sphericities, local fluctuations and also statistical biases. 
Moreover, numerous density peaks from substructures are also
apparent.   Notice that the best fit profile is given by the set of
parameters ($\alpha=0.39$, $\beta=3.72$,
 $\gamma=0.254$, and $a=13.16$ kpc), but its physical significance is arguable, given the large density fluctuations aforementioned.

As there is no baryonic component in this simulation, the sun
location can be chosen in any direction. 
For the full skymap pictures in the next section, two positions were chosen,
one along the positive $z$ axis (which is also the projection axis in
Figs. \ref{fig:simu} and
\ref{fig:clumpcircle}) and one along the positive $x$ axis.
The corresponding dark matter densities coincide to within 10\% and are  $\rho_0=0.0046\ {\rm M_{\odot}
 pc^{-3}}$ = 0.17 ${\rm GeVcm^{-3}}$ and $\rho_0=0.0043\ {\rm M_{\odot}
 pc^{-3}}$ = 0.165 ${\rm GeVcm^{-3}}$ respectively. Throughout the
rest of the paper, we will use the former for normalization purposes.

 The density around a simulation point $i$ was calculated with the
 algorithm of Casertano and Hut~\cite{1985ApJ29880C}, namely
 \begin{equation}
 \rho^i_j=\frac{j-1}{V(r_j)}M_p
 \end{equation}
 where $V(r_j)=4\pi/3 r_j^3$ is the volume of
 the smallest sphere around the particle $i$ that includes $j$ neighbors.
 Excluding the particle $i$ itself as well as the $j$th neighbor in the
 mass count gives an unbiased estimator of the density,
 with a variance $\sigma_j^2=\rho^2/(j-2)$. The choice of the number of
 neighbors used to calculate the density is a trade-off between reducing
 fluctuations and preserving
 the locality of the value computed by this method.
 A value $j\simeq 10$ seems to give a satisfactory compromise, as we could
 check on some Plummer test models.
 Higher values of $j$ wash out substructures and inhomogeneities present in
 the data, while smaller values of $j$ imply large statistical
 uncertainties that mask these inhomogeneities with Poissonian noise.
 As can be seen in Fig.~\ref{fig:rhoDM}, the density inside clumps can be
 several orders of magnitude higher than the density
 of the smooth component at that location.

 \subsection{Clumps}

 To identify clumps and subclumps in our simulation, we used the code
 ADAPTAHOP~\cite{Aubert:2004mu}, which is an improved algorithm based on HOP
 that enables to build a tree of structures and substructures.
 Basically, the algorithm divides the simulation points into disconnected
 groups, or leaves, corresponding to local density maxima. To decrease
 statistical noise, smoothing techniques are applied to calculate the
 density.
 The connections between leaves are created by performing a search of
 saddle points between groups.
 The density of a saddle point is then compared with the local maxima on
 each side, as well as a threshold parameter,
 to decide whether the structures are connected or not. By progressively
 raising this threshold from a minimum value
 corresponding to a galactic halo overdensity, and performing recursively
 the last check, the algorithm constructs a tree of (sub)structures. We
 note that the peak patches output by ADAPTAHOP are disconnected in space,
 as they are limited by the closest saddle points. As a consequence, clump masses
 given by this algorithm are often underestimated if other structures are present in their neighborhood.
 We found 108 (sub)clumps attached to the galaxy.
 Their spatial distribution is illustrated in
 Fig.~\ref{fig:clumpcircle}, where each clump is represented by a
 circle whose radius scales as the 1/3 power of its mass, $M^{1/3}$. A
 comparison with Fig.~\ref{fig:simu} shows that the two pictures are
 consistent. All the clumps within the virial radius are successfully
 identified by the algorithm. Outside this radius, some clumps are found as
 disconnected from the main galaxy, and therefore not visible on
 Fig.~\ref{fig:clumpcircle}.

  \begin{figure}[t]
 \begin{center}
 \begin{tabular}{ccc}
 \psfrag{fit}[ct][rb][1][1]{{\tiny $N(M_{sub}) \propto
 M_{sub}^{-1}$}}
 \includegraphics[width=0.32\textwidth]{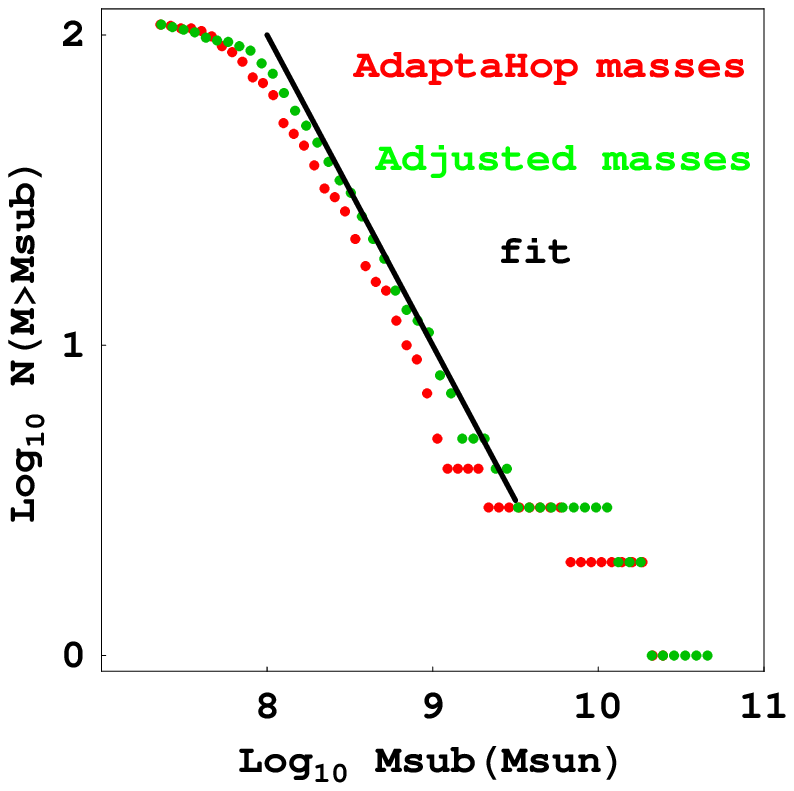}&
 \includegraphics[width=0.32\textwidth]{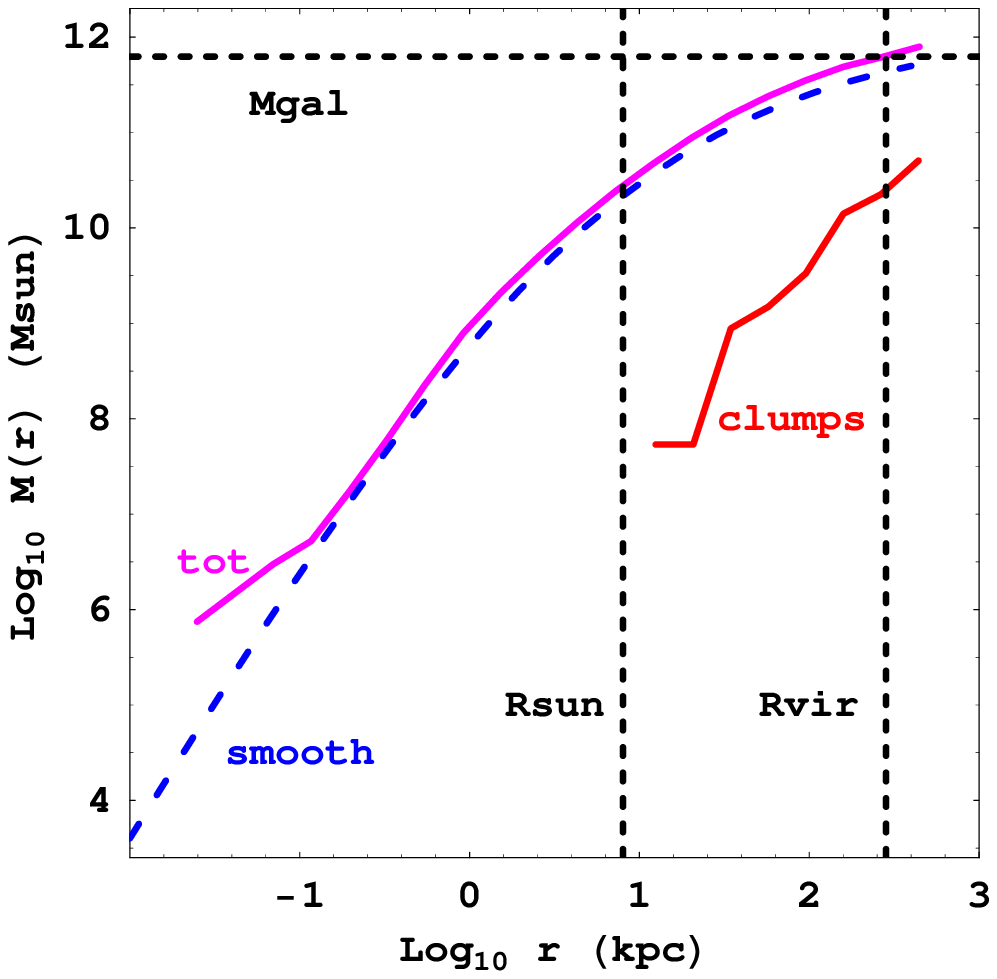}&
 \psfrag{fit}[c][c][1][1]{{\scriptsize $M\propto r^{-1.53}$}}
 \includegraphics[width=0.34\textwidth]{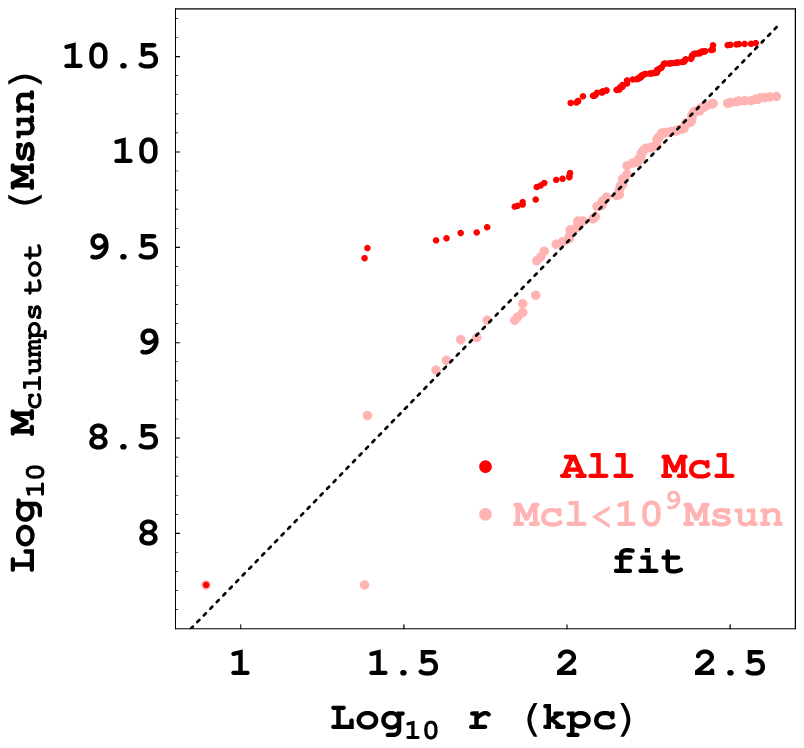}\\
 a) & b) & c)
 \end{tabular}
 \caption{\small a) Cumulative subhalo mass function of the clumps in the galaxy
   obtained with ADAPTAHOP masses (red - dark grey) and after the
   virial mass adjustment procedure (green - lighter grey). b) Cumulative mass for the simulation (tot), the
   smooth component and the clumps. c) A closer look at the cumulative mass profile for the clumps 
   (all clumps in red, and only clumps lighter than $10^9\ M_\odot$ in
   pink).}
 \label{fig:clumpspectrum}
 \end{center}
 \end{figure}
 
 From this, we derived a cumulative subhalo mass function shown on
 Fig.~\ref{fig:clumpspectrum}a. 
 The clumps mass fraction inside the virial radius is equal to $3.6\%$ of the total mass within
   that radius when taking the mass values output by ADAPTAHOP.
 We have corrected these masses by fitting a density profile on each clump,
 and then extrapolating it to the clump virial radius.
 The adjusted clumps mass fraction inside the virial radius is then equal to
 $5.4\%$.
 The mass adjustment has an impact on the mass function, as can be seen on
 Fig.~\ref{fig:clumpspectrum}a. The result is compatible with a power-law
 with index $-1$ above a mass
 threshold of 5 $\times 10^8$ $M_\odot$.
 For smaller masses, a flattening of the curve is apparent,
 due to the resolution limit of the simulation, which does not
 allow the survival of clumps with mass below a given threshold. This
 is around $200$ particles, in agreement with \cite{Diemand:2006ik}. 
The effect of the mass adjustment further enhances this flattening
 at low masses. Even after this correction, the final mass function
 that we get is still lower than that found
 by other authors \cite{Stoehr:2003hf,Diemand:2006ik}, but this deviation is not
 statistically significant.
For example, we found 3 (7) clumps with mass before (after)
adjustment higher than $10^9\ M_\odot$, compared to 13 in
 \cite{Diemand:2006ik}.

 For the clump radial number density distribution,  the statistics is too
 low to obtain reliable information about the behavior near the
 galactic center (GC).
 Instead, it is preferable to look at the cumulative mass profile as a
 function of the radius (distance from the GC), shown on
 Fig.~\ref{fig:clumpspectrum}b and c.
 When the clumps with a mass $> 10^9$ $M_\odot$ are removed, the cumulative
 mass profile is well fit by a power-law with index $n \simeq 1.75$ up to the virial radius.
The corresponding clump mass density profile with index $n-3 \simeq -1.25$ is thus flatter than the smooth
component. 
The mass fraction in the form of clumps is therefore increasing with
 radius.
This can be intuitively understood, since clump merging should be easier
 in the central parts.
However, the clump density in this simulation is too low to dominate the mass fraction
at any radius up to the virial radius.

Examples of density profiles within two clumps are given in
Fig.~\ref{fig:insideclumps}. Except for the inner parts, they 
are compatible with a power law with slope
$\simeq - 2.5$ (Fig.\ref{fig:insideclumps}). It is clear that further
in there is a sharp transition, but we do not have sufficient points
to constrain the inner slopes. In Fig.~\ref{fig:insideclumps} we have,
as an illustration, plotted two power laws with slopes 0 and 1,
respectively.   
The transition between the outer slope and the inner slope is described by a 
concentration parameter $c_{vir}=r_{vir}^{cl}/r_{-2}$, where $r_{-2}$ is the radius at which 
$d/dr\left(r^2\rho(r)\right)=0$ and $r_{vir}^{cl}$ is the virial radius of the clump.
Several models predict that the concentration parameter strongly
correlates with the virial mass (figure 2 of \cite{Pieri:2007ir}).

 \section{Gamma and neutrino fluxes from dark matter annihilation}

In this work, we consider that the dark matter particle candidate is a
typical WIMP provided by some new physics beyond the standard model. 
The gamma ray or neutrino flux per solid angle unit from the annihilation of dark matter particles (with mass $m_{DM}$,
density $\rho_{DM}$, cross-section $\langle \sigma v \rangle$, and branching ratios $BR_i$ into final state $i$) 
can be written as
 \begin{equation}
 \frac{d\Phi_{\gamma,\nu}}{d\Omega}= \frac{1}{4 \pi} \quad \underbrace{\frac{1}{\delta}
 \frac{\langle \sigma v \rangle}{ m_{DM}^2}
  \quad \int_{E^{\gamma,\nu}_{min}}^{E^{\gamma,\nu}_{max}}\sum_i \frac{dN^i_{\gamma,\nu}}{dE_{\gamma,\nu}}
  BR_i }_{\doteq HEP_{\gamma,\nu}} \quad
  \underbrace{ \int_{l(\vec{\Omega})}
 \rho_{DM}^2\, dl~~,}_{\doteq ASTRO}
 \label{flux}
 \end{equation}
where $dN^i_{\gamma,\nu}/dE_{\gamma,\nu}$ is the differential gamma/neutrino spectrum per annihilation 
coming from the decay of annihilation products of final state $i$,
the integral is taken along the line of sight with direction $\vec{\Omega}$,
and $\delta=2$ for a self conjugate dark matter particle and 4 otherwise.
We have separated into two brackets factors that arise from particle physics and
from astrophysics.

The annihilation signal is proportional to the density squared, and can therefore benefit from
a strong enhancement if the dark matter distribution is highly clumpy.
This enhancement is known as the {\it boost factor}.

 \subsection{Astrophysics factor}

To further discuss the enhancement due to the distribution, it is useful
to define the dimensionless quantity
\begin{equation}
\bar{J}(\Delta \Omega) = \frac{1}{\Delta \Omega} \frac{1} {\rho^2_0
 r_0} \int\,\rho^2\,\, d l \, d\Omega,
\label{eq:jbar}
\end{equation}
where the solid angle $\Delta \Omega$ can be taken as the experimental
solid angle resolution 
of a given experiment such as GLAST.

To evaluate this quantity for our simulation, two different methods can be used.
In the first method, $\rho^2$ can be calculated at any coordinate of
the simulation box space with
the Casertano-Hut algorithm. Note that the estimator of $\rho^2$ is smaller than the square
of the estimator for $\rho$ by a factor $(j-2)/(j-1)$. The integral
along a given line of sight
is then calculated  with the method of rectangles, with a variable step equal to half
of the distance to the $j=10$th closest neighbor. This ensures that the integral will not be
overestimated when local density peaks are encountered along the line of sight.
Finally, the value of $\bar{J}$ in a cone is the average of the values
for different lines of sight within that cone.
Fig.~\ref{fig:Jbarskymap} presents an all-sky view of the astrophysical factor $\bar{J}$ in a 
Hammer projection for a value $\Delta \Omega = 10^{-5}$ relevant for
GLAST. Since the simulation does not include a baryonic component, the
position of the observed is only constraied by its distance from the
center, i.e. can be anywhere on a sphere of radius 8 kpc. We
calculated the $\bar{J}$ all sky map for two different viewing
positions to allow comparisons. The shape of the iso-
$\bar{J}$ contours differs significantly between the two cases and is
not circular-like. Indeed cosmological
simulations show that dark matter halos are not spherical, but have
the shape of triaxial ellipsoids (e.g. \cite{BailinSteinmetz:2005,Allgood06}) and this is true also
for the simulation we analyze here. Thus, the observed signal will
depend on the viewing angle, since the integral in Eq.~(\ref{eq:jbar})
will have different values if the integration is e.g. along a major or
a minor axis of the ellipsoidal shape. This is also noted in the zoom
of the central region, shown in Fig.~\ref{fig:GLASTzoomsky}. In this
figure we also see that the relevant size of the central region is of
the order of a kpc, which has been found by galactic dynamic
simulations to have a rich structure in the baryonic component,
including inner bars, inner discs, rings and/or spirals. Whether these
influence in any way the dark matter in that region still remains to
be studied.


We also evaluated the astrophysical factor with a second method, in which the integral in Eq.~(\ref{eq:jbar})
is replaced by a finite sum over simulation points.
\begin{equation}
\hat{J}(\Delta \Omega) = \frac{1}{\Delta \Omega} \frac{1}{\rho_0^2 r_0}
\sum_{i \in \Delta \Omega}  \frac{\rho_i M_p}{l_i^2} 
\end{equation}
The presence of a pole in $l^{-2}$ is potentially dangerous in this method, and can lead to an overestimation of the fluxes
in case a simulation point is very close to the observer's location.
Nevertheless for this simulation and for the chosen sun locations,
we get comparable results for the two methods, although fluxes from low density
regions suffer more from statistical noise. 
To compare the two numerical methods more globally,
we calculated $\bar{J}$ and $\hat{J}$ for the whole sky, $\Delta \Omega=4\pi$.
We found  a good agreement ($\sim 2\%$) between the two
methods. 

 \begin{figure}
 \begin{center}
 \begin{tabular}{ccc}
 \includegraphics[width=0.33\textwidth]{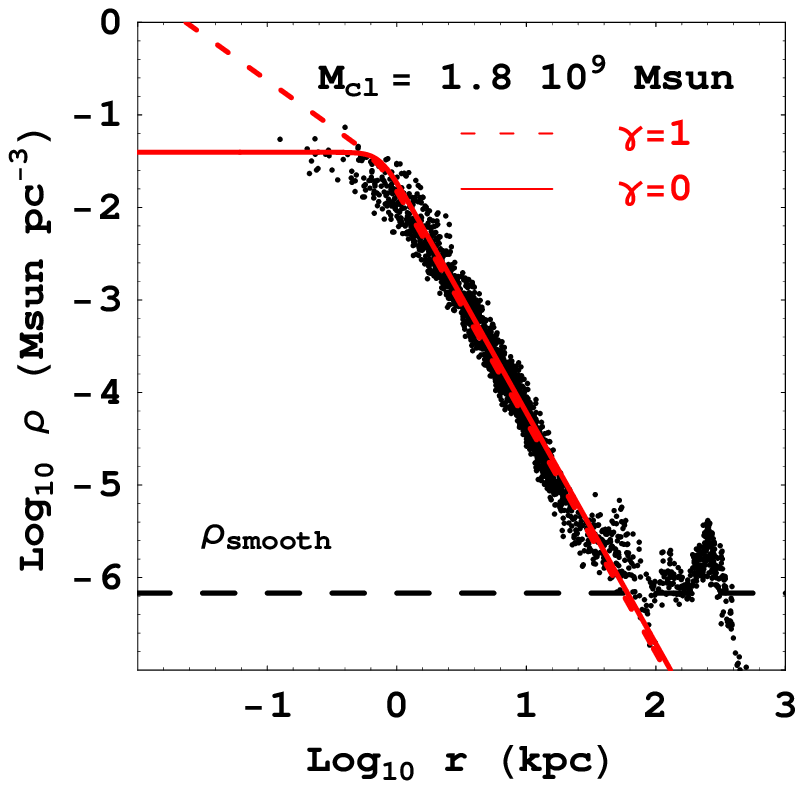}&
 \includegraphics[width=0.33\textwidth]{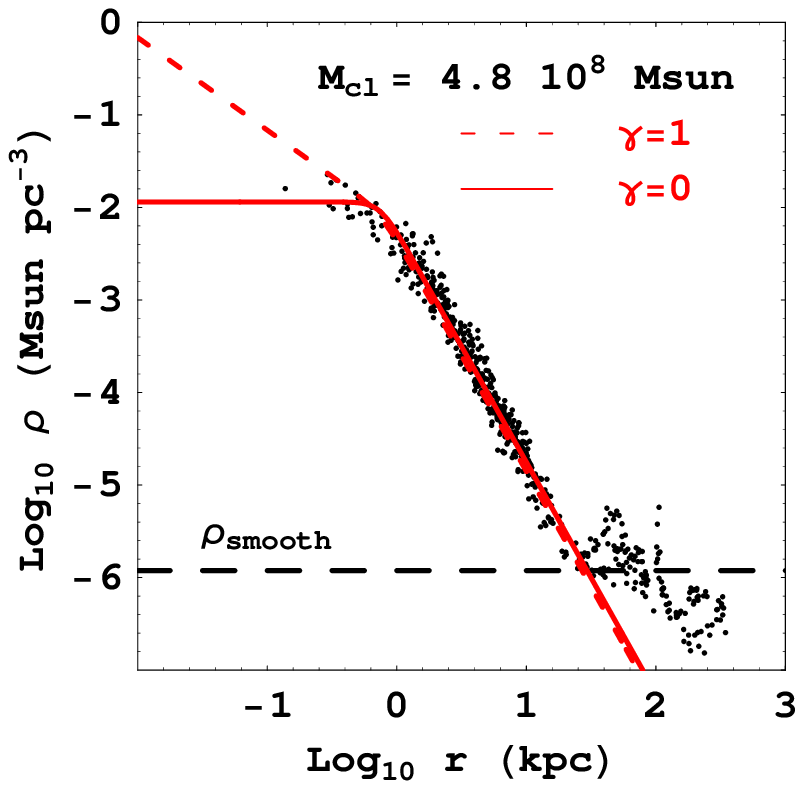}&
 \includegraphics[width=0.33\textwidth]{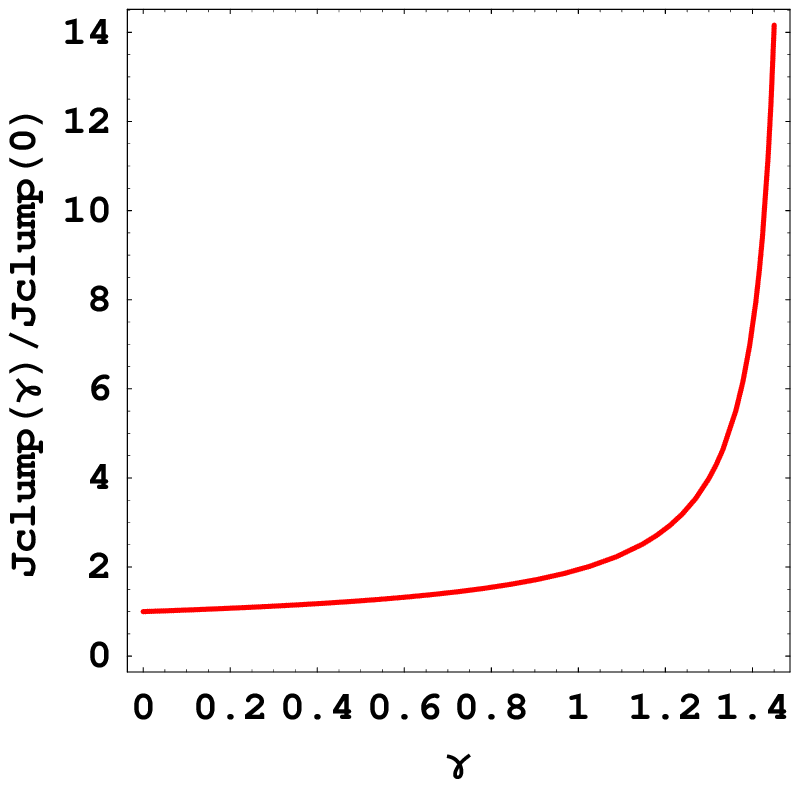}\\
 a) & b) &c)
 \end{tabular}
 \caption{\small a) and b) Density profile inside two clumps. The
   outer slope of the profile corresponds to $\beta=2.5$. c) Clump
   luminosity boost factor as a function of the exponent of the clump
   inner density profile $(\gamma)$.}
 \label{fig:insideclumps}
 \end{center}
\end{figure}

 \subsection{Particle Physics factor}

 \begin{figure}
 \begin{center}

 \includegraphics[angle=-90,width=0.95\textwidth]{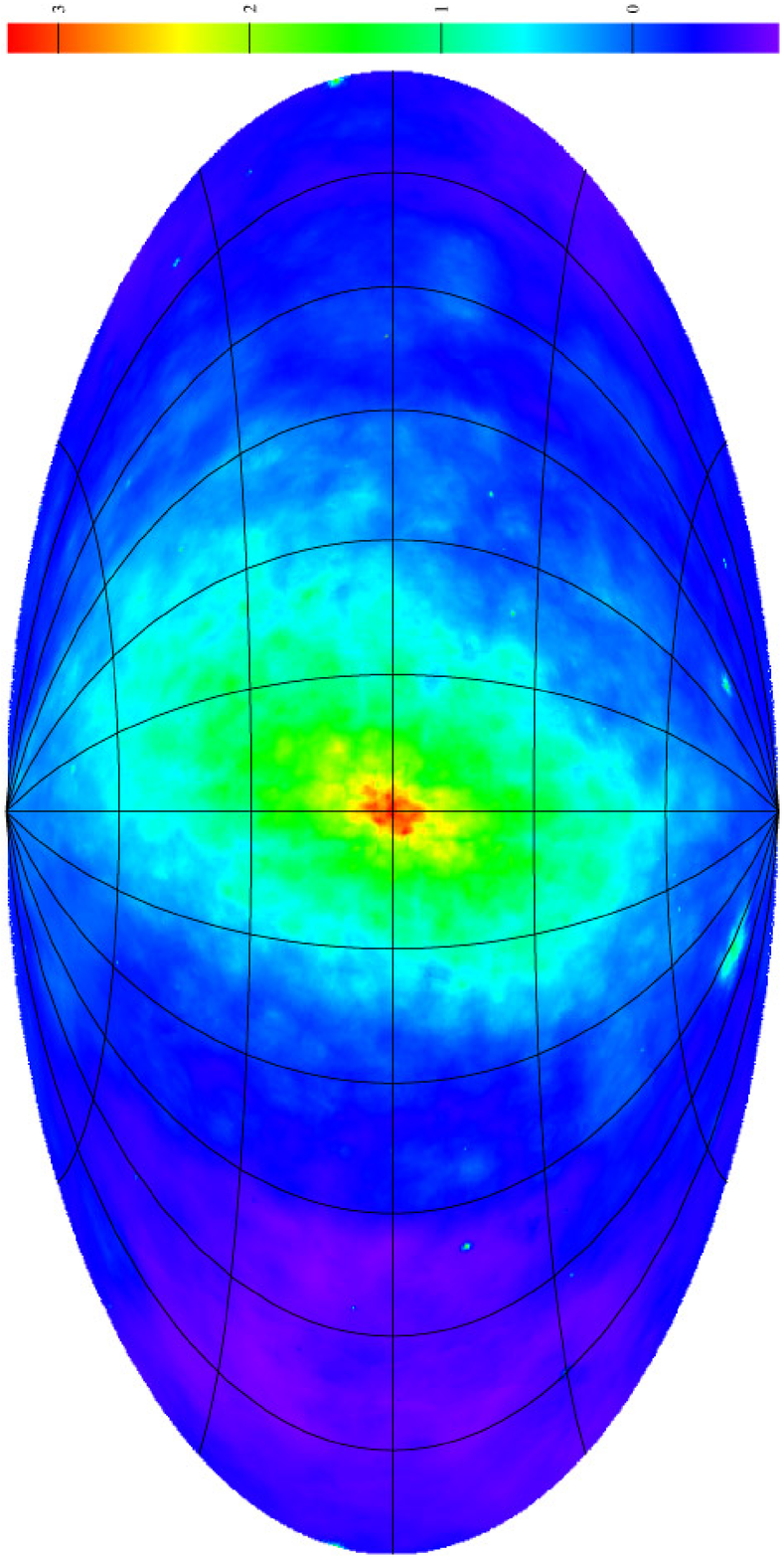}\\

 \includegraphics[angle=-90,width=0.99\textwidth]{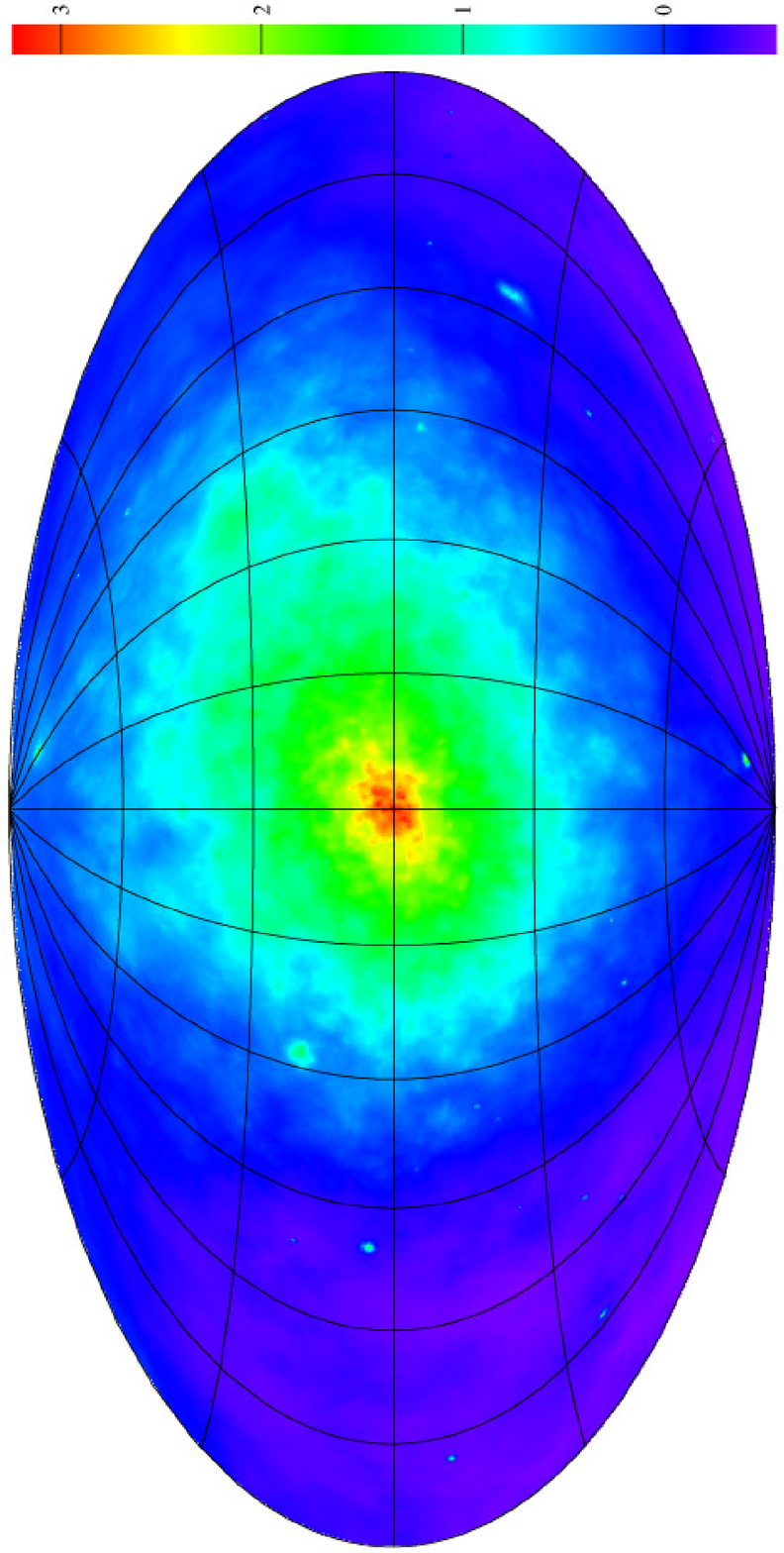}

 \caption{\small $\bar{J}$ all sky map in a Hammer projection. The observer positions are $(0,0,R_{sun})$  (top figure)  and
 $(R_{sun},0,0)$  (bottom figure). The values range from less than 1 in the anticenter to more than $10^3$ in the direction of the center.}
 \label{fig:Jbarskymap}
 \end{center}
 \end{figure}

The evaluation of the particle physics contribution ($HEP_{\gamma,\nu}$) in Eq.~(\ref{flux}) is highly dependent
on the physics beyond the standard model that one assumes. 
Let us rewrite the HEP term as 
\begin{equation}
HEP_{\gamma,\nu}=\frac{1}{\delta} \frac{\langle \sigma v \rangle}{ m_{DM}^2} N_{\gamma,\nu} \qquad,
\label{eq:hep}
\end{equation}
where $N_{\gamma,\nu}$ is the number of photons/neutrinos per  annihilation
 \begin{equation}
 N_{\gamma,\nu} =  \int_{E^{\gamma,\nu}_{min}}^{E^{\gamma,\nu}_{max}}\sum_i
 \frac{dN^i_{\gamma,\nu}}{dE_{\gamma,\nu}} BR_i \qquad.
 \end{equation}

The spectra coming from dark matter particles annihilation spread up to $m_{DM}$. 
The number of photons/neutrinos depends on the decay chain of the dark matter 
annihilation products and these annihilations are determined by the particular particle
 physics model considered. 
 We can estimate this number by using an effective and quite model independent
 approach. Namely, we will assume, as in typical BSM model, that dark matter
 particles (like those coming from SUSY, extra-dimendions ...) annihilate into SM
 (Standard Model) particles ($W^+W^-, \tau^+\tau^-, b\bar{b},t\bar{t} ...$)
 whose decays will , in turn, produce the gamma/neutrino continuum.
The spectrum can then be evaluated from Pythia~\cite{Sjostrand:2006za} simulations.

For gamma rays, the actual number of photons in a given experiment depends on its energy range.
For GLAST, the energy range will be $\approx 1-300$ GeV. These threshold values fix the actual limits for $E^{\gamma}_{min}$ 
and $E^{\gamma}_{max}$. 
In the 1-300 GeV GLAST energy range, the number of photons estimated from Pythia simulations~\cite{Mambrini:2005vk} typically yields $N_{\gamma}\sim$ 1, 10, 50 and
 200  for $m_{DM}\sim$ 10, 100, 1000 GeV and 10 TeV respectively.

For neutrinos, we will consider the ANTARES sensitivity and a possible improvement
for a km3 size telescope located in the mediterranean sea. The typical
energy threshold for such a neutrino telescope is $\sim$ 100 GeV
leading to $1\lesssim N_\nu\lesssim 10$  for  $100\ {\rm GeV}\lesssim
m_{DM}\lesssim 10^4\ {\rm GeV}$ (derived from Pythia neutrino spectrum~\cite{Labonne:2006hk}).

 The other important quantity in Eq.~(\ref{eq:hep}) is the annihilation rate $\langle \sigma
 v \rangle$. From the cosmological point of view, $\langle \sigma v
 \rangle$ can be
 related to the relic abundance of dark matter, roughly
 $\Omega_{DM}\sim 1/\langle \sigma v \rangle$, even if strictly speaking
 the temperature is higher in the primordial plasma at the freeze-out
 than in the surrounding sky at the present epoch and $\langle \sigma v
 \rangle_{T_0}\not=  \langle \sigma v
 \rangle_{freeze-out} $. For a standard annihilation scenario with thermal
 freeze-out, a value of $\Omega_{DM}$ like what was found by WMAP implies $\langle \sigma v
 \rangle\sim 10^{-26}\ {\rm cm^3s^{-1}}$. Of course, different hypotheses
 can lead to variations in this value. For instance, one can
 consider some models 
 where coannihilations drive the number of relic particles.  Moreover,
 cosmological scenarios with low reheating
 temperature could allow to decrease the annihilation cross section and
 still satisfy the WMAP constraint.
 Conversely, scenarios with dark matter particle production through
 late decays (out of equilibrium) of heavier particles could lead to a
 higher annihilation cross section to respect the WMAP relic density. Considering those
 sources of variability, almost independently of any specific particle physics
 framework, we will assume the reasonable range $ 10^{-27} \lesssim
 \langle \sigma v \rangle [ \mathrm{cm^3.s^{-1}}]\lesssim 10^{-24}$.

 Consequently, taking into account the production of photons and neutrinos and the
 annihilation rate uncertainties, a typical range for the HEP contribution
 to the gamma and neutrinos fluxes are given by

 \begin{equation}
 10^{-33}\quad \lesssim \quad HEP_{\gamma}\,\, [{\rm photons\,\,
 cm^3.s^{-1}.GeV^{-2}} ] \quad \lesssim \quad 10^{-26} \qquad,
 \end{equation}

 \begin{equation}
 10^{-34}\quad \lesssim \quad HEP_{\nu}\,\, [{\rm neutrinos\,\,
 cm^3.s^{-1}.GeV^{-2}} ] \quad \lesssim \quad 10^{-28} \qquad,
 \end{equation}
where the difference between gamma and neutrino is due to the
higher energy threshold in neutrino experiments, which reduces $N_\nu$ and
 leads to consider higher values of the dark matter mass~\footnote{This
  gives typically 2-3 orders of magnitude between EGRET and
  ANTARES performances despite their similar
  sensitivities ($\sim10^{-8-7}\gamma(\nu)\ {\rm cm^{-2}s^{-1}}$). The same is true for explicit SUSY models~\cite{Bertone:2004ag}.}.

 \subsection{Comparison with experiments }
 \label{section:glast}

 With typical values for the local dark matter density $\rho_0\simeq
 0.3\, {\rm GeVcm^{-3}}$ \cite{Kamionkowski:2008vw} and for the Sun to Galactic Center distance
 $r_0\simeq 8$ kpc, one has
 \begin{equation}
 \Phi_{\gamma,\nu}\, [{\rm photons/neutrinos \ cm^{-2}.s^{-1}}] \sim 10^{20}\, HEP_{\gamma,\nu}\,
 \bar{J}\ \Delta \Omega.
 \label{fluxlast}
 \end{equation}

\subsubsection{GLAST}
 GLAST \cite{Gehrels:1999ri} is a satellite which should be launched
 this year. Its angular resolution should be $\Delta \Omega \sim 10^{-5}$
 srad ({\it i.e.} an opening angle of 0.1 degrees ).  Considering the
 sky sensitivity of GLAST given by \cite{Bertone:2006kr}, we take $10^{-10}\, {\rm
 photons\,\, cm^{-2}s^{-1}}$ as a reasonable value to determine the interesting
 benchmark region in excess of which we could expect a signal to be
 detected by GLAST.

\begin{figure}

 \begin{center}
 \begin{tabular}{cc}
 \psfrag{kpc}[1][1]{ {\scriptsize kpc}}
 \includegraphics[width=0.49\textwidth]{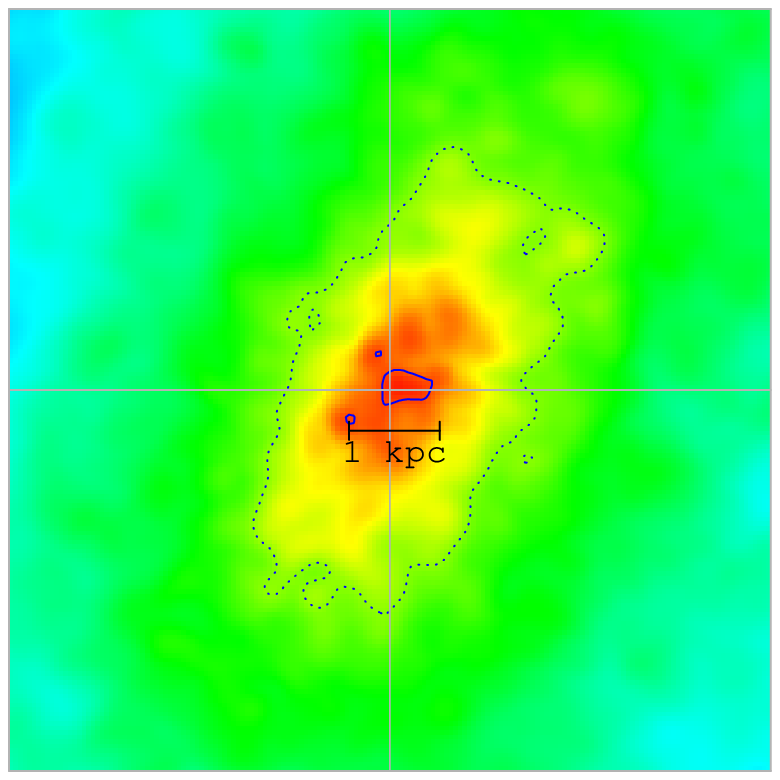}&
 \psfrag{kpc}[1][1]{ {\scriptsize kpc}}
 \includegraphics[width=0.49\textwidth]{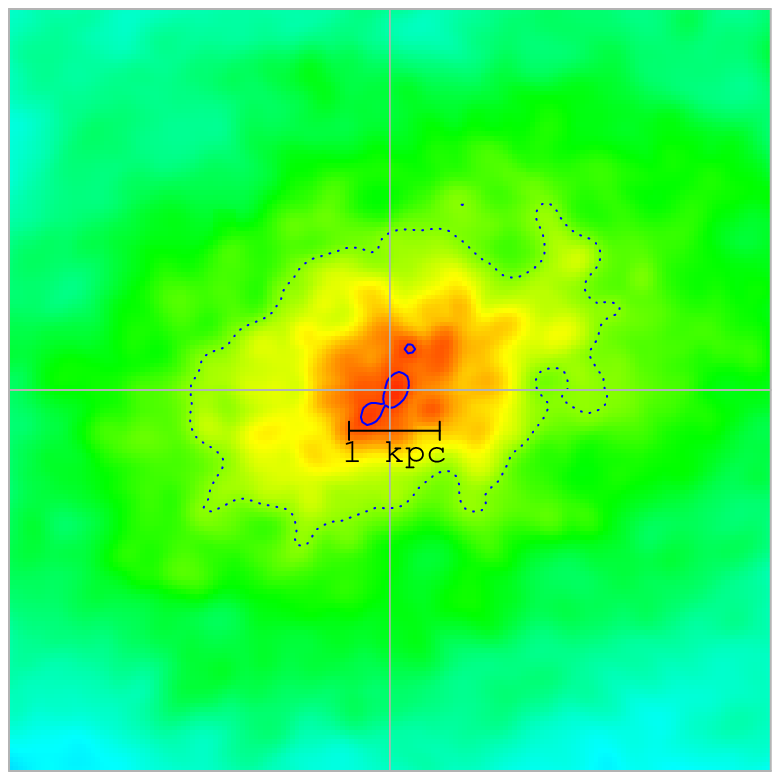}\\

 a) & b)
 \end{tabular}

 \caption{\small a) Zoom in the $ 60^{\circ} \times 60^{\circ}$ central
 region of Fig.~\ref{fig:Jbarskymap}, i.e. a region centered around
 the direction of the Galactic center. The horizontal bar gives the
 distance of 1 kpc, which is roughly the region which will be an
 interesting target for GLAST. The contours correspond to
 $\bar{J}=10^{3}$ (solid) and  $\bar{J}=10^{2}$ (dotted).
 b) Same as a) but with observer
 position at $(R_{sun},0,0)$ instead of $(0,0,R_{sun})$ in a). The
 two plots are normalized with the same local density value.}
 \label{fig:GLASTzoomsky}
 \end{center}
 \end{figure}

In Fig.~\ref{fig:GLASTzoomsky}, we show a zoom of the central region
of Fig.~\ref{fig:Jbarskymap} for the two viewing angles analyzed.
The image size is 60 deg $\times$ 60 deg. 
To smooth the artificial substructures that are due to noise in the simulation,
we averaged the fluxes in a 5x5=25 pixels square that corresponds to a linear resolution of $200$ pc.
Taking into account the $ HEP_\gamma$ contribution, we indicate
the $\bar{J}$ values of $10^{3}$ and $10^{2}$ 
as quite optimistic GLAST benchmarks.
The two figures show $\bar{J}$ values normalized with the {\it
same} local density. One has to keep in mind that the local density 
can well vary by an order of magnitude depending on the Sun
location (see Fig.~\ref{fig:rhoDM}) and that this will influence
accordingly the calculated flux value.
 
\subsubsection{ANTARES and a km3 size neutrino telescope}
AMANDA and ICECUBE are located at the south pole, so that the galactic
center region, which is the most promising one, is very challenging for
those experiments. Thus, we consider in this section the
ANTARES experiment, whose deployment completion is imminent, and extrapolate the sensitivity for a possible km3 size
neutrino telescope in the mediterranean sea like the KM3NeT
project~\cite{Carr:2007zc} associating ANTARES, NEMO and NESTOR collaborations. The resolution of Antares
depends on the neutrino-muon angle, but is typically  $\Delta \Omega
\sim 10^{-3}$ srad. We consider
the galactic center sensitivity of ANTARES for dark matter derived in
\cite{DavePhD}, the recent improvement of effective area due to trigger update
\cite{Lim:2007dg} and the expected performance for a km3 size telecope
\cite{Carr:2007zc}. Taking those references into account, we
believe that the sensitivity above 100 GeV of a future km3 size neutrino
telescope located in the mediterranean sea  should be around
 $10^{-9}\, {\rm
 neutrinos\,\, cm^{-2}s^{-1}}$ for point sources in the sky.
In Fig. \ref{fig:AntaEgretzoomsky} we show the central region
$\bar{J}$ skymap but calculated with a $\Delta \Omega= 10^{-3}$
resolution corresponding to the proposed km3 neutrino telescope resolution. The contour shows an optmistic
benchmark region corresponding to $\bar{J}=100$ in the zoom
for GLAST (no region is available for $\bar{J}=10^3$).

\subsubsection{EGRET}

\begin{figure}

 \begin{center}
 \begin{tabular}{cc}
 \psfrag{kpc}[1][1]{ {\scriptsize kpc}}
 \includegraphics[width=0.49\textwidth]{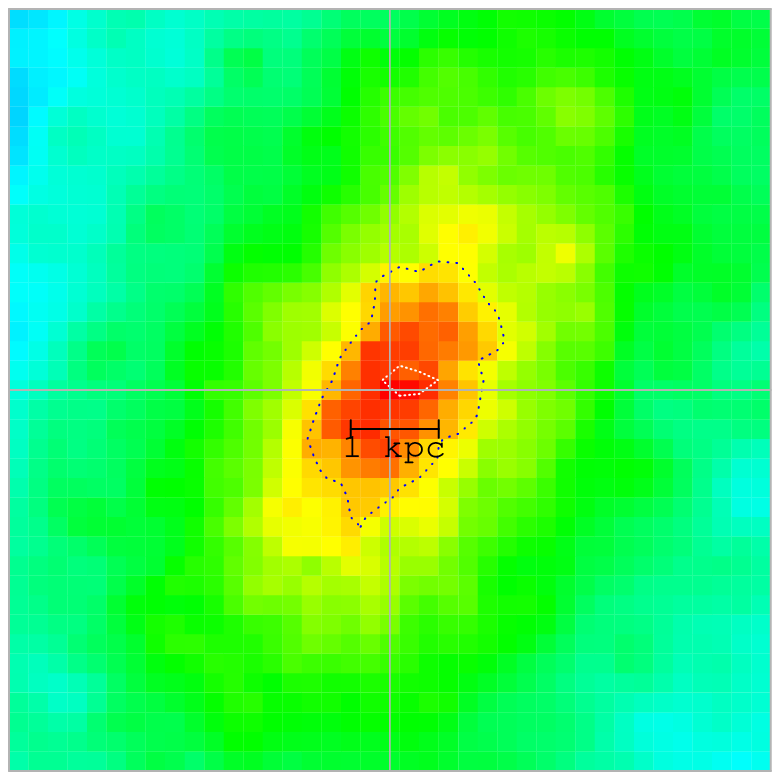}&
 \psfrag{kpc}[1][1]{ {\scriptsize kpc}}
 \includegraphics[width=0.49\textwidth]{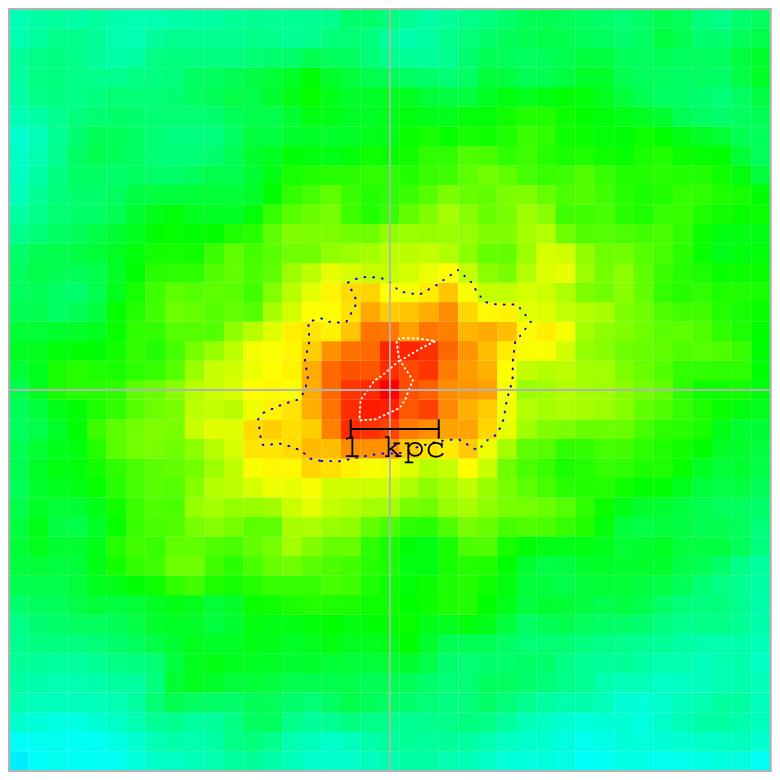}\\

 a) & b)
 \end{tabular}

 \caption{\small a) Same as figure \ref{fig:GLASTzoomsky} with $\Delta \Omega=
   10^{-3}$ for a km3 neutrino telescope
   and EGRET, i.e. the $ 60^{\circ} \times 60^{\circ}$
 region around the Galactic center of the $\bar{J}(\Delta \Omega
 =10^{-3})$ skymap. The light contour corresponds to an
 hypothetical km3 size neutrino telescope, the dark one corresponds to
 the level of the controversial EGRET excess.
 b) Same as a) but with observer
 position at $(R_{sun},0,0)$ instead of $(0,0,R_{sun})$ in a). The
 two plots are normalized with the same local density value.}
 \label{fig:AntaEgretzoomsky}
 \end{center}
 \end{figure}

A few years ago, the EGRET collaboration reported an excess in gamma ray fluxes above 1 GeV~\cite{egret}. 
After subtracting the cosmic ray background, a residual flux of around $5 \times 10^{-8}{\rm  photons\,\, cm^{-2}s^{-1}}$
remains. The excess is hard to accommodate with natural cosmic ray propagation models, as it requires
a harder electron injection spectrum or average spectrum in the galaxy different from the local ones~\cite{Strong:2004de}.
As an alternative, the annihilation of dark matter has been suggested.
To fit the energy spectrum, the dark matter candidate should have a mass around 50 GeV.
With an angular resolution $\Delta \Omega= 10^{-3}$ srad ({\it i.e.}
an opening angle of 0.1 degrees), the value of the residual flux would show as the contour on
Fig.~\ref{fig:AntaEgretzoomsky}, when taking an optimistic HEP$_\gamma$ that would correspond to $\bar{J}=10^2$ for GLAST. 
The central region would indeed be observable by Egret, while the regions at higher
longitudes or latitudes remain in the background.
However, this is not what was observed by Egret, as the excess is found to be rather constant with latitude or longitude,
except close to the galactic center region where it is mildly higher.
When fitted to the data, the hypothesis of
dark matter yields a structure which is incompatible with 
the NFW profile. Instead, an prolate isothermal halo supplemented by two rings gives good results~\cite{deBoer:2006tv}.
As a consequence, a constant boost factor of order 40 is needed in
this WIMP candidate scenario, and this implies 
excessive secondary antiprotons fluxes, in clear contradiction with observations~\cite{Bergstrom:2006tk}. 
Moreover, it has been recently argued that the EGRET excess might be due to a calibration problem rather than
any unknown astrophysical or exotic contribution~\cite{Stecker:2007xp}.

\subsection{Sensitivity to clumps and resolution}

Taken at face value, our results seem to argue that 
no clump in the simulation seems to be within the reach of GLAST.
The intrinsic luminosity of an individual clump is determined by its mass and its concentration,
but the corresponding flux is damped by the distance square factor.
Inside the virial radius, the clumps contribute for $8.2\%$ of the total luminosity for a mass fraction
of $5.4\%$, but only for $0.2\%$ of the total flux. This, however,
depends strongly on the distance from the nearest clump. A further
point to take into account is the resolution limit of the simulation.  
Due to that, the innermost behavior of the clumps profile in poorly
determined, as already mentioned.
The intrinsic luminosity of a clump can be boosted by increasing the inner slope $\gamma$.
In Fig.~\ref{fig:insideclumps}c, we show the clump luminosity boost factor as a function of $\gamma$, 
at constant mass and concentration. For $\gamma$ values less than 1.3
or 1.4 the boost factor is moderate. In
particular, the boost for $\gamma=1$ (NFW profile) compared to
$\gamma=0~(core)$ 
is only around a factor of $2$. However, for values of $\gamma$ above
1.3 the boost factor increases very spectacularly, to reach a a factor
of 10 around $\gamma=1.4$. 

To have a much stronger effect, the concentration parameter should be increased.   
If we take a NFW profile for the clump, it is easy to check that the concentration parameter 
is directly expressed in terms of the scale radius $a$ and the virial radius of the clump $r_{cl}$ as $ c_{vir}=r_{cl}/a$.
Then the total annihilation $\bar{J}$ resulting from a clump of virial mass $M_{vir}^{cl}$ sitting at a distance $l \gg r_{cl}$
will be given by 
 \begin{equation}
J_{cl} \simeq \frac{4\pi}{3} \frac{r_{cl}^2}{l^2} \frac{r_{cl}}{r_0} \frac{\rho_{vir}^2}{\rho_0^2} f_{cl}(c_{vir}),
 \end{equation}
where $\rho_{vir}$ is the virial density (200 times the cosmological matter density), and
 \begin{equation}
f(x) = \frac{(1+x^2)^2}{x} \left(1-\frac{1}{(1+x)^3}\right) \simeq x^3
 \end{equation}
for $x \gg 1$. The clump mass and clump size are related by the following relation
 \begin{equation}
M_{cl} = 4\pi \rho_{vir} r_{cl}^3 \left(\ln (1+c_{vir})- \frac{c_{vir}}{(1+c_{vir})} \right)
 \end{equation}
From this -- as $\rho_{vir}/\rho_0 \simeq 10^{-3}$ and since the
probability of having a clump in our immediate neighborhood is low --
we see that small (in angular size) clumps will be 
visible by GLAST only if they are highly concentrated, $c_{vir} \sim 10^2-10^3$. 

 \section{ Conclusion \& Perspectives}

We evaluated the gamma and neutrino fluxes from dark matter annihilation in
a galactic halo framework extracted from a cosmological N-body
simulation of the HORIZON project. Although such simulations are the most elaborate 
and realistic framework for this kind of studies,  there are very few
works concerning dark matter detection in N-body frameworks.
With reasonable assumptions on the new physics beyond the
standard model and on the cosmological scenario for dark matter particles, we
proposed an absolute evaluation of gamma and neutrino fluxes. This allowed to test the galactic
simulation framework with regard to dark matter indirect
detection in future experiments like GLAST and the proposed km3
size neutrino telescope in the mediterranean sea extrapolated from
ANTARES sensitivity and preliminary KM3NeT studies.

Our framework and even other cosmological simulations have a resolution limit which is lower than
what would be desirable for this type of studies. Nevertheless, it was possible to reach a number of interesting
results in our framework. In particular, we showed that the galactic center
region is a good benchmark part  of the sky with regard to the
GLAST sensitivity. Even if it is more challenging, this region should 
also be studied by future neutrino telescopes, especially a km3 size telescope. 
Individual clumps stand out clearer in the
direction of the galactic anticenter. The concentration parameters of
the clumps are then the crucial information that will determine the flux.
Unfortunately, our simulation does not provide enough information on scales below 100
pc to clearly conclude about the detectability of individual subhalos.  
Furthermore, it should be stressed that our results for the central
region of the Galaxy and for the clumps should be considered 
as a lower bound. A resolution improvement will increase the values of
the central density and decrease the size of the central region within
which we have little or no information. This will increase the signal
in the direction of the Galactic center, as well as in the directions
of the clumps. 
We can hope that specific geometry, non sphericity and structures
especially in the central region will be highlighted by future gamma 
and neutrino observations.

Cosmological simulations made very important progress in the last few
years, and a study of gamma-ray induced by dark matter
annihilation in the most precise N-BODY framework can be found in
\cite{Kuhlen:2007wv} with a resolution allowing the identification
of $\lesssim 10^6 \ M_{sun}$ clumps. Though the quality of this work
is impressive, this
minimal scale is still more than 10 orders of magnitude above the typical WIMP
free-streaming scale ($\sim 10^{-12-4} M_{sun}$ \cite{2006PhRvL..97c1301P}) and also
considerably bigger than the minimal surviving clump mass, which
anyway is still under debate.

\section*{Acknowledgments}
We would like to thank V. Bertin, A. Bosma and J. Lavalle for helpful discussions.
FSL work is supported by a FNRS belgium grant and the IAP. EN work
was also partially supported by a FNRS grant.
We also acknowledge the french Astroparticle Program and the grant
ANR-06-BLAN-0172 for financial support.

 \bibliographystyle{unsrt}
 \bibliography{simu-gamneutID}

\begin{thebibliography}{10}

\bibitem{Spergel:2006hy}
D.~N. Spergel et~al.
\newblock Wilkinson microwave anisotropy probe (wmap) three year results:
  Implications for cosmology.
\newblock 0300.

\bibitem{Bosma:2003yv}
Albert Bosma.
\newblock Dark matter in galaxies: Observational overview.
\newblock {\em Dark Matter in Galaxies, IAU symposium series,}, 220, 2004.

\bibitem{White:1987yr}
Simon D.~M. White, Carlos~S. Frenk, Marc Davis, and George Efstathiou.
\newblock Clusters, filaments, and voids in a universe dominated by cold dark
  matter.
\newblock {\em Astrophys. J.}, 313:505--516, 1987.

\bibitem{Clowe:2003tk}
Douglas Clowe, Anthony Gonzalez, and Maxim Markevitch.
\newblock Weak lensing mass reconstruction of the interacting cluster
  1e0657-558: Direct evidence for the existence of dark matter.
\newblock {\em Astrophys. J.}, 604:596--603, 2004.

\bibitem{Ferreras:2007na}
Ignacio Ferreras, Prasenjit Saha, and Scott Burles.
\newblock Unveiling dark halos in lensing galaxies.
\newblock 2007.

\bibitem{Jungman:1995df}
Gerard Jungman, Marc Kamionkowski, and Kim Griest.
\newblock Supersymmetric dark matter.
\newblock {\em Phys. Rept.}, 267:195--373, 1996.

\bibitem{Bertone:2004pz}
Gianfranco Bertone, Dan Hooper, and Joseph Silk.
\newblock Particle dark matter: Evidence, candidates and constraints.
\newblock {\em Phys. Rept.}, 405:279--390, 2005.

\bibitem{Athanassoula:2003yw}
E.~Athanassoula.
\newblock Bars and the connection between dark and visible matter.
\newblock 2003.

\bibitem{Allgood:2006}
Brandon Allgood, Ricardo~A. Flores, Joel~R. Primack, Andrey~V. Kravtsov,
  Risa~H. Wechsler, Andreas Faltenbacher, and James~S. Bullock.
\newblock The shape of dark matter haloes: dependence on mass, redshift, radius
  and formation.
\newblock {\em Mon. Not. Roy. Astron. Soc.}, 367:1781--1796, 2006.

\bibitem{Athanassoula:2007ex}
E.~Athanassoula.
\newblock A bar in the inner halo of barred galaxies i. structure and
  kinematics of a representative model.
\newblock {\em Mon. Not. Roy. Astron. Soc.}, 377:1569--1578, 2007.

\bibitem{Heller:2007tr}
Clayton Heller, Isaac Shlosman, and E.~Athanassoula.
\newblock Structure formation inside triaxial dark matter halos: Galactic
  disks, bulges and bars.
\newblock 2007.

\bibitem{Navarro:1996gj}
Julio~F. Navarro, Carlos~S. Frenk, and Simon D.~M. White.
\newblock A universal density profile from hierarchical clustering.
\newblock {\em Astrophys. J.}, 490:493--508, 1997.

\bibitem{Kravtsov:1997dp}
Andrey~V. Kravtsov, Anatoly~A. Klypin, James~S. Bullock, and Joel~R. Primack.
\newblock The cores of dark matter dominated galaxies: theory vs. observations.
\newblock {\em Astrophys. J.}, 502:48, 1998.

\bibitem{Moore:1999gc}
Ben Moore, Tom Quinn, Fabio Governato, Joachim Stadel, and George Lake.
\newblock Cold collapse and the core catastrophe.
\newblock {\em Mon. Not. Roy. Astron. Soc.}, 310:1147--1152, 1999.

\bibitem{Stoehr:2003hf}
Felix Stoehr, Simon D.~M. White, Volker Springel, Giuseppe Tormen, and Naoki
  Yoshida.
\newblock Dark matter annihilation in the milky way's halo.
\newblock {\em Mon. Not. Roy. Astron. Soc.}, 345:1313, 2003.

\bibitem{Diemand:2006ik}
Jurg Diemand, Michael Kuhlen, and Piero Madau.
\newblock Dark matter substructure and gamma-ray annihilation in the milky way
  halo.
\newblock {\em Astrophys. J.}, 657:262, 2007.

\bibitem{Kuhlen:2007wv}
Michael Kuhlen, Jurg Diemand, and Piero Madau.
\newblock Glast and dark matter substructure in the milky way.
\newblock {\em AIP Conf. Proc.}, 921:135--138, 2007.

\bibitem{Kuhlen:2008aw}
Michael Kuhlen, Jurg Diemand, and Piero Madau.
\newblock {The Dark Matter Annihilation Signal from Galactic Substructure:
  Predictions for GLAST}.
\newblock 2008.

\bibitem{Tasitsiomi:2002vh}
Argyro Tasitsiomi and A.~V. Olinto.
\newblock The detectability of neutralino clumps via atmospheric cherenkov
  telescopes.
\newblock {\em Phys. Rev.}, D66:083006, 2002.

\bibitem{Pieri:2003cq}
Lidia Pieri and Enzo Branchini.
\newblock On dark matter annihilation in the local group.
\newblock {\em Phys. Rev.}, D69:043512, 2004.

\bibitem{Koushiappas:2003bn}
Savvas~M. Koushiappas, Andrew~R. Zentner, and Terrence~P. Walker.
\newblock The observability of gamma-rays from neutralino annihilations in
  milky way substructure.
\newblock {\em Phys. Rev.}, D69:043501, 2004.

\bibitem{Bi:2005im}
Xiao-Jun Bi.
\newblock Gamma rays from the neutralino dark matter annihilations in the milky
  way substructures.
\newblock {\em Nucl. Phys.}, B741:83--107, 2006.

\bibitem{Pieri:2007ir}
L.~Pieri, G.~Bertone, and E~Branchini.
\newblock Dark matter annihilation in substructures revised.
\newblock 2007.

\bibitem{Evans:2003sc}
N.~W. Evans, F.~Ferrer, and Subir Sarkar.
\newblock {A 'Baedecker' for the dark matter annihilation signal}.
\newblock {\em Phys. Rev.}, D69:123501, 2004.

\bibitem{Pieri:2005pg}
Lidia Pieri, Enzo Branchini, and Stefan Hofmann.
\newblock Difficulty of detecting minihalos via gamm rays from dark matter
  annihilation.
\newblock {\em Phys. Rev. Lett.}, 95:211301, 2005.

\bibitem{Oda:2005nv}
Takeshi Oda, Tomonori Totani, and Masahiro Nagashima.
\newblock {Gamma-ray background from neutralino annihilation in the first
  cosmological objects}.
\newblock {\em Astrophys. J.}, 633:L65--L68, 2005.

\bibitem{Koushiappas:2006qq}
Savvas~M. Koushiappas.
\newblock Proper motion of gamma-rays from microhalo sources.
\newblock {\em Phys. Rev. Lett.}, 97:191301, 2006.

\bibitem{Baltz:2006sv}
Edward~A. Baltz, James~E. Taylor, and Lawrence~L. Wai.
\newblock Can astrophysical gamma ray sources mimic dark matter annihilation in
  galactic satellites?
\newblock 2006.

\bibitem{Strigari:2006rd}
Louis~E. Strigari, Savvas~M. Koushiappas, James~S. Bullock, and Manoj
  Kaplinghat.
\newblock {Precise constraints on the dark matter content of Milky Way dwarf
  galaxies for gamma-ray experiments}.
\newblock {\em Phys. Rev.}, D75:083526, 2007.

\bibitem{Strigari:2007at}
Louis~E. Strigari et~al.
\newblock {The Most Dark Matter Dominated Galaxies: Predicted Gamma- ray
  Signals from the Faintest Milky Way Dwarfs}.
\newblock 2007.

\bibitem{horizonweb}
The horizon project, http://www.projet-horizon.fr.

\bibitem{Ramses}
R.~Teyssier.
\newblock Cosmological hydrodynamics with adaptive mesh refinement. a new high
  resolution code called ramses.
\newblock {\em A\&A}, 385:337--364, 2002.

\bibitem{1985ApJ29880C}
S.~{Casertano} and P.~{Hut}.
\newblock {Core radius and density measurements in N-body experiments
  Connections with theoretical and observational definitions}.
\newblock {\em APJ}, 298:80--94, November 1985.

\bibitem{Aubert:2004mu}
Dominique Aubert, Christophe Pichon, and Stephane Colombi.
\newblock The origin and implications of dark matter anisotropic cosmic infall
  on ~l* haloes.
\newblock {\em Mon. Not. Roy. Astron. Soc.}, 352:376, 2004.

\bibitem{BailinSteinmetz:2005}
Matthias Bailin, Jeremy;~Steinmetz.
\newblock {Internal and External Alignment of the Shapes and Angular Momenta of
  LCDM Halos}.
\newblock {\em Astrophys. J.}, 627:647--665, 2005.

\bibitem{Allgood06}
B.~et~al. Allgood.
\newblock {The shape of dark matter haloes: dependence on mass, redshift,
  radius and formation}.
\newblock {\em MNRAS}, 367:1781--1796, 2006.

\bibitem{Sjostrand:2006za}
Torbjorn Sjostrand, Stephen Mrenna, and Peter Skands.
\newblock {PYTHIA 6.4 physics and manual}.
\newblock {\em JHEP}, 05:026, 2006.

\bibitem{Mambrini:2005vk}
Y.~Mambrini, C.~Munoz, E.~Nezri, and F.~Prada.
\newblock Adiabatic compression and indirect detection of supersymmetric dark
  matter.
\newblock {\em JCAP}, 0601:010, 2006.

\bibitem{Labonne:2006hk}
Benjamin Labonne, Emmanuel Nezri, and Jean Orloff.
\newblock {The suppression of neutralino annihilation into Z h}.
\newblock {\em Eur. Phys. J.}, C47:805--814, 2006.

\bibitem{Bertone:2004ag}
Gianfranco Bertone, Emmanuel Nezri, Jean Orloff, and Joseph Silk.
\newblock {Neutrinos from dark matter annihilations at the galactic centre}.
\newblock {\em Phys. Rev.}, D70:063503, 2004.

\bibitem{Kamionkowski:2008vw}
Marc Kamionkowski and Savvas~M. Koushiappas.
\newblock Galactic substructure and direct detection of dark matter.
\newblock arXiv:0801.3269 [astro-ph].

\bibitem{Gehrels:1999ri}
N.~Gehrels and P.~Michelson.
\newblock Glast: The next-generation high energy gamma-ray astronomy mission.
\newblock {\em Astropart. Phys.}, 11:277--282, 1999.

\bibitem{Bertone:2006kr}
Gianfranco Bertone, Torsten Bringmann, Riccardo Rando, Giovanni Busetto, and
  Aldo Morselli.
\newblock Glast sensitivity to point sources of dark matter annihilation.
\newblock astro-ph/0612387.

\bibitem{Carr:2007zc}
J.~Carr et~al.
\newblock {Sensitivity studies for the cubic-kilometre deep-sea neutrino
  telescope KM3NeT}.
\newblock Proceedings of 30th International Cosmic Ray Conference (ICRC 2007),
  Merida, Yucatan, Mexico, 3-11 Jul 2007. arXiv:0711.2145 [astro-ph].

\bibitem{DavePhD}
D.~Bailey.
\newblock {\em Monte Carlo tools and analysis methods for understanding the
  ANTARES experiment and predicting its sensitivity to Dark Matter}.
\newblock Ph.D. thesis 2002,
  http://antares.in2p3.fr/Publications/thesis/2002/d-bailey-thesis.ps.gz.

\bibitem{Lim:2007dg}
Gordon Lim.
\newblock {Indirect search for Dark Matter with the ANTARES neutrino
  telescope}.
\newblock Proceedings of 15th International Conference on Supersymmetry and the
  Unification of Fundamental Interactions (SUSY07), Karlsruhe, Germany, 26 Jul
  - 1 Aug 2007. arXiv:0710.3685 [astro-ph].

\bibitem{egret}
S.D. et~al. Hunter.
\newblock Egret observations of the diffuse gamma ray emission from the
  galactic plane.
\newblock {\em The Astrophysical Journal}, 481(1):205--240, 1997.

\bibitem{Strong:2004de}
Andrew~W. Strong, Igor~V. Moskalenko, and Olaf Reimer.
\newblock {Diffuse Galactic continuum gamma rays. A model compatible with EGRET
  data and cosmic-ray measurements}.
\newblock {\em Astrophys. J.}, 613:962--976, 2004.

\bibitem{deBoer:2006tv}
Wim de~Boer, C.~Sander, V.~Zhukov, A.~V. Gladyshev, and D.~I. Kazakov.
\newblock {EGRET excess of diffuse galactic gamma rays interpreted as a signal
  of dark matter annihilation}.
\newblock {\em Phys. Rev. Lett.}, 95:209001, 2005.

\bibitem{Bergstrom:2006tk}
L.~Bergstrom, J.~Edsjo, Michael Gustafsson, and P.~Salati.
\newblock {Is the dark matter interpretation of the EGRET gamma excess
  compatible with antiproton measurements?}
\newblock {\em JCAP}, 0605:006, 2006.

\bibitem{Stecker:2007xp}
F.~W. Stecker, S.~D. Hunter, and D.~A. Kniffen.
\newblock {The Likely Cause of the EGRET GeV Anomaly and its Implications}.
\newblock {\em Astropart. Phys.}, 29:25--29, 2008.

\bibitem{2006PhRvL..97c1301P}
S.~{Profumo}, K.~{Sigurdson}, and M.~{Kamionkowski}.
\newblock {What Mass Are the Smallest Protohalos?}
\newblock {\em Physical Review Letters}, 97(3):031301, July 2006.

\end{thebibliography}
 \end{document}